\documentstyle[prd,aps,preprint,tighten,epsfig]{revtex}

\begin{document}

\draft

\title{Seesaw Invariance of Fritzsch-like Neutrino Mass Matrices, 
Leptogenesis and Lepton Flavor Violation}
\author{{\bf Zhi-zhong Xing} ~ and ~ {\bf Shun Zhou}} 
\address{CCAST (World Laboratory), P.O. Box 8730, Beijing 100080, China \\
and Institute of High Energy Physics, Chinese Academy of Sciences, \\
P.O. Box 918 (4), Beijing 100049, China \\
({\it Electronic address: xingzz@mail.ihep.ac.cn}) }
\maketitle

\begin{abstract}
We propose a simple ansatz for neutrino phenomenology,
in which the relevant lepton mass matrices take the universal
Fritzsch-like form and the seesaw relation holds
under a particular condition. There exist six
textures of this nature, but their consequences
on neutrino oscillations are exactly the same.
We show that our scenario is viable to account for the 
cosmological baryon number asymmetry via thermal leptogenesis.
Its predictions for the lepton-flavor-violating decays 
$\mu \rightarrow e\gamma$, $\tau \rightarrow \mu\gamma$ and 
$\tau \rightarrow e\gamma$ are also presented. We find that 
the branching ratios of these rare processes depend strongly 
upon the phase parameters responsible for leptogenesis and 
for leptonic CP violation in neutrino oscillations.
\end{abstract}


\newpage

\section{Introduction}

The Super-Kamiokande \cite{SK}, SNO \cite{SNO}, KamLAND \cite{KM} 
and K2K \cite{K2K} neutrino oscillation experiments have 
provided us with very compelling evidence that neutrinos are
massive and lepton flavors are mixed. This important result 
indicates that the minimal supersymmetric standard model (MSSM)
is actually as incomplete as the standard model in describing
the neutrino phenomenology. A very simple but natural extension of 
the MSSM is to include one right-handed neutrino in 
each of three lepton families, while the Lagrangian of electroweak 
interactions keeps invariant under $\rm SU(2)_L \times U(1)_Y$ 
gauge transformation. In this case, the Lagrangian responsible for 
lepton masses can be written as
\begin{equation}
- {\cal L}_{\rm lepton} \; = \; 
\overline{l}_{\rm L} Y_l E H_1 + 
\overline{l}_{\rm L} Y_\nu N H_2 + 
\frac{1}{2} \overline{N^{\rm c}} M_{\rm R} N + {\rm h.c.} \; , 
\end{equation}
where $l_{\rm L}$ denotes the left-handed lepton doublet, 
$E$ and $N$ stand respectively for the 
right-handed charged lepton and Majorana neutrino singlets, and 
$H_1$ and $H_2$ (with hypercharges $\pm 1/2$) are the MSSM Higgs
doublets. After spontaneous gauge symmetry 
breaking, one obtains the charged lepton mass matrix 
$M_l \equiv Y_l \langle H_1\rangle$ and the Dirac neutrino mass matrix 
$M_{\rm D} \equiv Y_\nu \langle H_2\rangle$.
The scale of $M_{\rm R}$ may 
be considerably higher than $\langle H_{1,2}\rangle$, because right-handed 
neutrinos are $\rm SU(2)_L$ singlets and their mass term is not subject 
to electroweak symmetry breaking. Thus the effective 
neutrino mass matrix $M_\nu$ can be derived from
$M_{\rm D}$ and $M_{\rm R}$  via the seesaw relation
$M_\nu = M_{\rm D} M^{-1}_{\rm R} M^T_{\rm D}$ \cite{SS}. Although this
elegant relation can qualitatively attribute the smallness of left-handed 
neutrino masses to the largeness of right-handed neutrino masses, it
is unable to make any concrete predictions unless a specific lepton 
flavor structure is assumed. Hence an appropriate combination of the 
seesaw mechanism and possible flavor symmetries or texture 
zeros \cite{Review} is practically needed, in order to quantitatively 
account for the neutrino mass spectrum and lepton flavor mixing.

One purpose of this paper is to incorporate the seesaw mechanism with
the Fritzsch-like textures of lepton mass matrices listed in Table 1. 
Those six patterns of $M_l$ and $M_\nu$ are actually isomeric \cite{XZ};
i.e., they have the same phenomenological consequences, although 
their structures are apparently different from one another. If 
$M_{\rm D}$ and $M_{\rm R}$ take the same Fritzsch-like form as
$M_l$ and $M_\nu$ do, then the seesaw relation   
$M_\nu = M_{\rm D} M^{-1}_{\rm R} M^T_{\rm D}$ will in general
be violated. We shall show that this relation can keep unchanged, 
provided the condition 
${\cal B}_{\rm D}/{\cal C}_{\rm D} = {\cal B}_{\rm R}/{\cal C}_{\rm R}$ 
is satisfied. In this case, each nonvanishing matrix element of 
$M_\nu$ has a simple seesaw relation with its
counterparts in $M_{\rm D}$ and $M_{\rm R}$; i.e.,
\begin{equation}
{\cal A}_\nu = \frac{{\cal A}^2_{\rm D}}{{\cal A}_{\rm R}} \; , ~~~~
{\cal B}_\nu = \frac{{\cal B}^2_{\rm D}}{{\cal B}_{\rm R}} \; , ~~~~
{\cal C}_\nu = \frac{{\cal C}^2_{\rm D}}{{\cal C}_{\rm R}} \; .
\end{equation}
The textures of $M_{\rm D}$, $M_{\rm R}$ and $M_\nu$ in Table 1 are
therefore referred to as the seesaw-invariant textures. Because
the parameters of $M_\nu$ can essentially be determined from current 
neutrino oscillation data \cite{X02}, it is then possible to impose 
useful constraints on the parameters of $M_{\rm D}$ and $M_{\rm R}$ 
via Eq. (2).

Another purpose of this paper is to account for the cosmological 
baryon number asymmetry via thermal leptogenesis \cite{FY}. Indeed,
lepton number violation induced by the third term of 
${\cal L}_{\rm lepton}$ allows decays of the heavy Majorana 
neutrinos $N_i$ (for $i=1,2,3$) to happen. Since the decay can occur 
at both tree and one-loop levels, their interference may result in a 
$CP$-violating asymmetry $\varepsilon_i$ between the $CP$-conjugated 
$N_i \rightarrow l + H^{\rm c}_2$ and $N_i \rightarrow l^{\rm c} + H_2$ 
processes. If the masses of $N_i$ are hierarchical 
(i.e., $M_1 < M_2 < M_3$), the interactions of $N_1$ can be in thermal 
equilibrium when $N_2$ and $N_3$ decay. The asymmetries $\varepsilon_2$ 
and $\varepsilon_3$ are therefore erased before $N_1$ decays, and only
the asymmetry $\varepsilon_1$ produced by the out-of-equilibrium decay
of $N_1$ survives. The point of leptogenesis \cite{FY} is that 
$\varepsilon_1$ may give rise to a net lepton number asymmetry 
$Y_{\rm L} \equiv (n^{~}_{\rm L} - n^{~}_{\rm\bar L})/{\bf s} 
\propto \varepsilon_1$, which is eventually converted into a net baryon 
number asymmetry $Y_{\rm B}$ via nonperturbative sphaleron 
processes \cite{Kuzmin}:
$Y_{\rm B} \equiv (n^{~}_{\rm B} - n^{~}_{\rm\bar B})/{\bf s} 
\propto Y_{\rm L}$. Thus this mechanism provides a natural interpretation 
of the cosmological matter-antimatter asymmetry, 
$7 \times 10^{-11} \lesssim Y_{\rm B} \lesssim 10^{-10}$, which
is drawn from the recent WMAP observational data \cite{WMAP}.
We shall show that six Fritzsch-like textures of lepton mass matrices 
in Table 1 yield the same $\varepsilon_1$ and $Y_{\rm B}$, from which
useful constraints on the mass scale of three right-handed neutrinos 
and the Majorana phase of CP violation can be obtained.

The third purpose of this paper is to calculate the rare
lepton-flavor-violating processes $\mu \rightarrow e \gamma$,
$\tau \rightarrow \mu \gamma$ and $\tau \rightarrow e \gamma$.
We shall focus our attention on a rather conservative case of
lepton flavor violation, based on the scenarios where
supersymmetry is broken in a hidden sector and the breaking
is transmitted to the observable sector by a flavor-blind mechanism. 
We find that the values of ${\rm Br}(l_j \rightarrow l_i \gamma)$ 
depend strongly upon the phase parameters responsible for leptogenesis 
and for CP violation in neutrino oscillations.

\section{Seesaw-invariant textures}

Now let us show that the seesaw relation 
$M_\nu = M_{\rm D} M^{-1}_{\rm R} M^T_{\rm D}$ will hold for each of
the six patterns listed in Table 1, if and only
if the condition 
${\cal B}_{\rm D}/{\cal C}_{\rm D} = {\cal B}_{\rm R}/{\cal C}_{\rm R}$ 
is satisfied. To be explicit, we take pattern (A) -- namely, the Fritzsch 
texture \cite{F78}, for example. Given the Fritzsch form of $M_{\rm D}$ 
and $M_{\rm R}$, the seesaw relation leads straightforwardly to
\begin{equation}
M_\nu \; = \; \left ( \matrix{
{\bf 0} & ~ \displaystyle\frac{{\cal C}^2_{\rm D}}{{\cal C}_{\rm R}} ~ 
& {\bf 0} \cr\cr
\displaystyle\frac{{\cal C}^2_{\rm D}}{{\cal C}_{\rm R}} & {\bf 0} & 
\displaystyle\frac{{\cal B}_{\rm D} {\cal C}_{\rm D}}
{{\cal C}_{\rm R}} \cr\cr
{\bf 0} & \displaystyle\frac{{\cal B}_{\rm D} {\cal C}_{\rm D}}
{{\cal C}_{\rm R}} & 
\displaystyle\frac{{\cal A}^2_{\rm D}}{{\cal A}_{\rm R}} \cr} \right ) +
\frac{{\cal C}^2_{\rm D}}{{\cal A}_{\rm R}} 
\left ( \frac{{\cal B}_{\rm D}}{{\cal C}_{\rm D}}
- \frac{{\cal B}_{\rm R}}{{\cal C}_{\rm R}} \right ) \left ( \matrix{
{\bf 0} & {\bf 0} & {\bf 0} \cr\cr
{\bf 0} & \displaystyle \frac{{\cal B}_{\rm D}}{{\cal C}_{\rm D}}
- \frac{{\cal B}_{\rm R}}{{\cal C}_{\rm R}} &
\displaystyle \frac{{\cal A}_{\rm D}}{{\cal C}_{\rm D}} \cr\cr
{\bf 0} & \displaystyle \frac{{\cal A}_{\rm D}}{{\cal C}_{\rm D}} 
& {\bf 0} \cr}
\right ) \; .
\end{equation}
One can see that the second term of Eq. (3) will vanish, if 
${\cal B}_{\rm D}/{\cal C}_{\rm D} = {\cal B}_{\rm R}/{\cal C}_{\rm R}$ 
is taken. In this case, the (2,3) or (3,2) element of $M_\nu$ 
turns out to be
${\cal B}_{\rm D}{\cal C}_{\rm D}/{\cal C}_{\rm R} = 
{\cal B}^2_{\rm D}/{\cal B}_{\rm R}$. Then Eq. (3) is simplified to
\begin{equation}
M_\nu \; =\; \left ( \matrix{
{\bf 0} & ~ \displaystyle\frac{{\cal C}^2_{\rm D}}
{{\cal C}_{\rm R}} ~ & {\bf 0} \cr\cr
\displaystyle\frac{{\cal C}^2_{\rm D}}{{\cal C}_{\rm R}} & {\bf 0} & 
\displaystyle\frac{{\cal B}^2_{\rm D}}{{\cal B}_{\rm R}} \cr\cr
{\bf 0} & \displaystyle\frac{{\cal B}^2_{\rm D}}{{\cal B}_{\rm R}} & 
\displaystyle\frac{{\cal A}^2_{\rm D}}{{\cal A}_{\rm R}} \cr} \right ) \; .
\end{equation}
Comparing between the form of $M_\nu$ in Eq. (4) and that defined
in Table 1, we immediately arrive at the elegant seesaw relations
given in Eq. (2). This result implies that 
\begin{equation}
\frac{{\cal B}_{\rm D}}{{\cal C}_{\rm D}} \; =\; 
\frac{{\cal B}_{\rm R}}{{\cal C}_{\rm R}} \; =\; 
\frac{{\cal B}_\nu}{{\cal C}_\nu} 
\end{equation}
holds. Eq. (5) imposes a strong constraint on the structure
of $M_{\rm D}$, $M_{\rm R}$ and $M_\nu$. Because the magnitudes of
${\cal A}_\nu$, ${\cal B}_\nu$ and ${\cal C}_\nu$ 
only have a quite weak hierarchy as
required by current experimental data \cite{XZ}, we are not 
allowed to assume 
${\cal B}_\nu/{\cal C}_\nu = {\cal B}_l/{\cal C}_l$ in a similar way. 
Indeed, $|{\cal B}_\nu|/|{\cal C}_\nu| \ll |{\cal B}_l|/|{\cal C}_l|$ 
must hold. Without loss of generality, we take ${\cal A}_l$, 
${\cal A}_\nu$, ${\cal A}_{\rm D}$ and ${\cal A}_{\rm R}$ to be real 
and positive. Then only the off-diagonal
elements of $M_a$ (for $a=l$, D, R or $\nu$) are complex.
It is possible to decompose $M_a$ into 
$M_a = P_a \overline{M}_a P_a^T$, where
\begin{equation}
\overline{M}_a \; =\; \left ( \matrix{
{\bf 0}         & C_a    & {\bf 0} \cr
C_a   & {\bf 0}       & B_a \cr
{\bf 0}         & B_a         & A_a \cr} \right ) \; 
\end{equation}
and $P_a = {\rm Diag}
\{e^{i(\varphi_a - \phi^{~}_a)}, ~e^{i\phi^{~}_a}, ~1\}$
with $A_a = {\cal A}_a$, $B_a = |{\cal B}_a|$, $C_a = |{\cal C}_a|$,
$\phi_a \equiv \arg ({\cal B}_a)$ and $\varphi_a \equiv \arg ({\cal C}_a)$.
Eq. (2) indicates that 
$\phi^{~}_\nu = 2 \phi^{~}_{\rm D} - \phi^{~}_{\rm R}$ and
$\varphi^{~}_\nu = 2 \varphi^{~}_{\rm D} - \varphi^{~}_{\rm R}$ hold.
Hence one can get the phase relation
$\phi^{~}_{\rm D} - \varphi^{~}_{\rm D} =
\phi^{~}_{\rm R} - \varphi^{~}_{\rm R} = 
\phi^{~}_\nu - \varphi^{~}_\nu$ from Eq. (5). Because of
${\rm Det}(\overline{M}_a) = -A_a C_a^2 <0$, it is more convenient to 
diagonalize $\overline{M}_a$ by using the transformation
\begin{equation} 
~~~ \left (O_aQ \right )^T \overline{M}_a \left (O_a Q \right ) \; = \; 
\left ( \matrix{
\lambda^a_1 & 0 & 0 \cr
0 & \lambda^a_2 & 0 \cr
0 & 0 & \lambda^a_3 \cr} \right ) \; ,
\end{equation}
where $O_a$ denotes a {\it real} orthogonal matrix, 
$Q = {\rm Diag}\{1, i, 1\}$ is a pure phase matrix defined to cancel 
the minus sign of ${\rm Det}(\overline{M}_a)$, and $\lambda^a_i$ 
(for $i=1,2,3$) stand for the {\it positive} eigenvalues of 
$\overline{M}_a$. Then we have
\begin{eqnarray}
A_a & = & \lambda^a_1 - \lambda^a_2 + \lambda^a_3 \; ,
\nonumber \\
B_a & = & \left [ \frac{\left (\lambda^a_1 - \lambda^a_2 \right ) 
\left (\lambda^a_2 - \lambda^a_3 \right ) \left (\lambda^a_1 + 
\lambda^a_3 \right )}
{\lambda^a_1 - \lambda^a_2 + \lambda^a_3} \right ]^{1/2}  ,
\nonumber \\
C_a & = & \left ( \frac{\lambda^a_1 \lambda^a_2 \lambda^a_3}
{\lambda^a_1 - \lambda^a_2 + \lambda^a_3} \right )^{1/2}  .
\end{eqnarray} 
Defining the dimensionless ratios 
$x_a \equiv \lambda^a_1/\lambda^a_2$ and 
$z_a \equiv \lambda^a_1/\lambda^a_3$, we further obtain
\begin{eqnarray}
O^a_{11} & = & + \left [ \frac{x_a - z_a}
{\left (1+x_a \right ) \left (1-z_a \right )
\left (x_a - z_a + x_a z_a \right )}  \right ]^{1/2} \; , 
\nonumber \\
O^a_{12} & = & - \left [ \frac{x^3_a \left (1+z_a \right )}
{\left (1+x_a \right ) \left (x_a + z_a \right ) 
\left (x_a - z_a + x_a z_a \right )} \right ]^{1/2} \; ,
\nonumber \\
O^a_{13} & = & + \left [ \frac{z^3_a \left ( 1 - x_a \right )}
{\left (1-z_a \right ) \left (x_a + z_a \right ) 
\left ( x_a - z_a + x_a z_a \right )} \right ]^{1/2} \; , 
\nonumber \\
O^a_{21} & = & + \left [ \frac{x_a - z_a}
{\left (1+x_a \right ) \left (1-z_a \right )} \right ]^{1/2} \; ,
\nonumber \\
O^a_{22} & = & + \left [ \frac{ x_a \left (1+z_a
\right )}{\left (1+x_a \right ) \left (x_a + 
z_a \right )} \right ]^{1/2} \; , 
\nonumber \\
O^a_{23} & = & + \left [ \frac{z_a \left (1-x_a \right )}
{\left (1-z_a \right ) \left (x_a + z_a \right )} \right ]^{1/2} \; ,
\nonumber \\
O^a_{31} & = & - \left [ \frac{x_a z_a \left (1-x_a \right ) 
\left (1+z_a \right )}{\left (1+x_a \right ) \left (1-z_a \right )
\left (x_a - z_a + x_a z_a \right )} \right ]^{1/2} \; , 
\nonumber \\
O^a_{32} & = & - \left [ \frac{z_a \left (1 -x_a \right ) 
\left (x_a - z_a \right )}{\left (1+x_a \right ) \left (x_a + z_a \right ) 
\left (x_a - z_a + x_a z_a \right )} \right ]^{1/2} \; , 
\nonumber \\
O^a_{33} & = & + \left [ \frac{x_a \left (1 +z_a \right ) 
\left (x_a - z_a \right )}{\left (1-z_a \right ) \left (x_a + 
z_a \right ) \left (x_a - z_a + x_a z_a \right )} \right ]^{1/2} \; .
\end{eqnarray}
The full calculability of $M_a$ is quite encouraging, because it implies
that the parameters of $M_{\rm D}$ can be determined in terms of
those of $M_\nu$ and $M_{\rm R}$:
\begin{equation}
A_{\rm D} = \sqrt{A_\nu A_{\rm R}} \; , ~~~
B_{\rm D} = \sqrt{B_\nu B_{\rm R}} \; , ~~~
C_{\rm D} = \sqrt{C_\nu C_{\rm R}} \; .
\end{equation}
It is then possible to link the observables of leptogenesis and
lepton flavor violation to the masses of three light neutrinos
($m_i$) and three heavy neutrinos ($M_i$) in a rather simple way.

One may straightforwardly show that the other five patterns of neutrino 
mass matrices in Table 1 are also seesaw-invariant under the condition
${\cal B}_{\rm D}/{\cal C}_{\rm D} = {\cal B}_{\rm R}/{\cal C}_{\rm R}$.
Thus Eqs. (2), (5) and (10) are universally valid. We diagonalize $M_a$
via the transformation
\begin{equation}
\left (P^*_a O_a Q \right )^TM_a \left (P^*_a O_a Q \right )
\; = \; \left (O_aQ \right )^T \overline{M}_a \left (O_a Q \right ) \; ,
\end{equation}
as defined in Eq. (7). The relevant forms of $P_a$ and $O_a$ 
are listed in Table 1 for six Fritzsch-like textures of lepton mass
matrices. Then we find that Eqs. (8) and (9) universally hold. This
result allows us to discuss six isomeric patterns of $M_a$ in a 
uniform way. 

The phenomenon of lepton flavor mixing arises from the mismatch between 
diagonalizations of $M_l$ and $M_\nu$. It is described by 
the unitary matrix $V = (P^*_l O_l Q)^T (P^*_\nu O_\nu Q)^*$. 
Explicitly,
\begin{equation}
|V_{pq}| \; = \; \left | O^l_{1p} O^\nu_{1q} e^{i\alpha} ~ + ~
O^l_{2p} O^\nu_{2q} e^{i\beta} ~ + ~ O^l_{3p} O^\nu_{3q} \right | \; ,
\end{equation}
where the subscripts $p$ and $q$ run respectively over 
$(e, \mu, \tau)$ and $(1,2,3)$, and the phases $\alpha$
and $\beta$ are defined by 
$\alpha \equiv (\varphi^{~}_\nu - \varphi^{~}_l) - \beta$ and
$\beta \equiv (\phi^{~}_\nu - \phi^{~}_l)$. It is clear that
$|V_{pq}|$ depend only upon four free parameters: $x_\nu$, $z_\nu$,
$\alpha$ and $\beta$, because the mass ratios of charged leptons
$x^{~}_l$ and $z^{~}_l$ are well known. The parameter space of
$(x_\nu, z_\nu)$ and $(\alpha, \beta)$ can be determined from 
current experimental data on solar, atmospheric and
reactor neutrino oscillations. A detailed analysis has been done in 
Ref. \cite{XZ}. It is found that the Fritzsch-like textures of
$M_l$ and $M_\nu$ can fit the present data at the $3\sigma$ level.

Instead of repeating the analysis done before, we shall subsequently
concentrate on the consequences of our phenomenological ansatz on
thermal leptogenesis and lepton flavor violation. 

\section{Thermal Leptogenesis}

As argued above, we assume that three heavy right-handed neutrinos
have a clear mass hierarchy and the out-of-equilibrium decay
of the lightest one ($N_1$) is the only source of lepton number
asymmetry. In the flavor basis where both $M_l$ and $M_{\rm R}$ are 
diagonal, real and positive, the CP-violating asymmetry between 
$N_1 \rightarrow l + H^{\rm c}_2$ and $N_1 \rightarrow l^{\rm c} + H_2$
processes reads \cite{ZZX04}
\begin{eqnarray}
\varepsilon_1 & \equiv & \frac{\Gamma (N_1 \rightarrow l + H^{\rm c}_2)
- \Gamma (N_1 \rightarrow l^{\rm c} + H_2)}
 {\Gamma (N_1 \rightarrow l + H^{\rm c}_2)
+ \Gamma (N_1 \rightarrow l^{\rm c} + H_2)} 
\nonumber \\
& \approx & -\frac{3}{8\pi} \cdot 
\frac{x^{~}_{\rm R} {\rm Im} \left [ (\tilde{M}_{\rm D}^\dagger 
\tilde{M}_{\rm D})_{12} \right ]^2 + z^{~}_{\rm R} {\rm Im} 
\left [ (\tilde{M}_{\rm D}^\dagger \tilde{M}_{\rm D})_{13} \right ]^2}
{\langle H_2\rangle^2 
(\tilde{M}_{\rm D}^\dagger \tilde{M}_{\rm D})_{11}} \; ,
\end{eqnarray}
in which $x^{~}_{\rm R} \equiv M_1/M_2$ and $z^{~}_{\rm R} \equiv M_1/M_3$
with a normal mass hierarchy $z^2_{\rm R} \ll x^2_{\rm R} \ll 1$,
$\langle H_2\rangle = v \sin\beta_{\rm susy}$ with $v \approx 174$ GeV, 
and $\tilde{M}_{\rm D} =(P^*_l O_l Q)^T M_{\rm D} 
(P^*_{\rm R} O_{\rm R} Q)$.
Note that $\tilde{M}_{\rm D}^\dagger \tilde{M}_{\rm D}$ can be 
expressed as
\begin{equation}
\tilde{M}_{\rm D}^\dagger \tilde{M}_{\rm D} \; =\;
\left (P O_{\rm R} Q \right )^\dagger 
\overline{M}^2_{\rm D} \left (P O_{\rm R} Q \right ) \; ,
\end{equation}
where 
$P \equiv P^T_{\rm D} P^*_{\rm R} = {\rm Diag}\{1, e^{i\sigma}, 1\}$
with $\sigma \equiv \phi^{~}_{\rm D} - \phi^{~}_{\rm R}$. In
obtaining this result, we have used the phase relation 
$\phi^{~}_{\rm D} - \varphi^{~}_{\rm D} =
\phi^{~}_{\rm R} - \varphi^{~}_{\rm R}$.
It is remarkable that only a single phase parameter $\sigma$ contributes
to $\tilde{M}_{\rm D}^\dagger \tilde{M}_{\rm D}$. If $\sigma$ vanishes,
there will be no CP violation in the lepton-number-violating decays
of heavy Majorana neutrinos $N_i$. We emphasize that
$\sigma$ has no direct connection with the effect of leptonic CP
violation in neutrino oscillations. The latter is actually associated
with the phase differences $\alpha$ and $\beta$ appearing in 
Eq. (12) \cite{X02}. Only in the special case that $\phi^{~}_l$ and 
$\varphi^{~}_l$ are switched off (or fixed to certain values), it is 
possible to indirectly link $\sigma$ to $\alpha$ and $\beta$. For example, 
$\phi^{~}_l = \varphi^{~}_l = 0$ leads to 
$\beta = \sigma + \phi^{~}_{\rm D}$ and 
$\alpha = \varphi^{~}_\nu - \beta$. 

With the help of Eqs. (6) and (14), we obtain the explicit expressions
of $(\tilde{M}^\dagger_{\rm D} \tilde{M}_{\rm D})_{11}$,
${\rm Im}[(\tilde{M}^\dagger_{\rm D} \tilde{M}_{\rm D})_{12}]^2$ and
${\rm Im}[(\tilde{M}^\dagger_{\rm D} \tilde{M}_{\rm D})_{13}]^2$
as follows:
\begin{eqnarray}
(\tilde{M}^\dagger_{\rm D} \tilde{M}_{\rm D})_{11} & = & 
A^2_{\rm D} (O^{\rm R}_{31})^2  
+ B_{\rm D}^2 \left [(O^{\rm R}_{21})^2 + (O^{\rm R}_{31})^2 \right ]
+ C_{\rm D}^2 \left [(O^{\rm R}_{11})^2 + (O^{\rm R}_{21})^2 \right ]
\nonumber \\
& & + 2 A_{\rm D} B_{\rm D} O^{\rm R}_{21} O^{\rm R}_{31} \cos\sigma 
+ 2 B_{\rm D} C_{\rm D} O^{\rm R}_{11} O^{\rm R}_{31} \; ,
\end{eqnarray}
and
\begin{eqnarray}
{\rm Im}[(\tilde{M}^\dagger_{\rm D} \tilde{M}_{\rm D})_{12}]^2 & = &
-2 A_{\rm D} B_{\rm D} \left ( O^{\rm R}_{22} O^{\rm R}_{31} 
- O^{\rm R}_{21} O^{\rm R}_{32} \right ) 
\left [ A^2_{\rm D} O^{\rm R}_{31} O^{\rm R}_{32} + 
B_{\rm D}^2 \left ( O^{\rm R}_{21} O^{\rm R}_{22} + 
O^{\rm R}_{31} O^{\rm R}_{32} \right ) \right .
\nonumber \\
& & \left . + C_{\rm D}^2 \left ( O^{\rm R}_{11} O^{\rm R}_{12} 
+ O^{\rm R}_{21} O^{\rm R}_{22} \right ) + 
A_{\rm D} B_{\rm D} \left ( O^{\rm R}_{21} O^{\rm R}_{32} +
O^{\rm R}_{22} O^{\rm R}_{31} \right ) \cos\sigma \right .
\nonumber \\
& & \left . + B_{\rm D} C_{\rm D} \left ( O^{\rm R}_{11} O^{\rm R}_{32} + 
O^{\rm R}_{12} O^{\rm R}_{31} \right ) \right ] \sin\sigma \; ,
\nonumber \\
{\rm Im}[(\tilde{M}^\dagger_{\rm D} \tilde{M}_{\rm D})_{13}]^2 & = &
+2 A_{\rm D} B_{\rm D} \left ( O^{\rm R}_{23} O^{\rm R}_{31} 
- O^{\rm R}_{21} O^{\rm R}_{33} \right ) 
\left [ A^2_{\rm D} O^{\rm R}_{31} O^{\rm R}_{33} + B_{\rm D}^2 
\left ( O^{\rm R}_{21} O^{\rm R}_{23} + 
O^{\rm R}_{31} O^{\rm R}_{33} \right ) \right .
\nonumber \\
& & \left . + C_{\rm D}^2 \left ( O^{\rm R}_{11} O^{\rm R}_{13} 
+ O^{\rm R}_{21} O^{\rm R}_{23} \right ) + 
A_{\rm D} B_{\rm D} \left ( O^{\rm R}_{21} O^{\rm R}_{33} +
O^{\rm R}_{23} O^{\rm R}_{31} \right ) \cos\sigma \right .
\nonumber \\
& & \left . + B_{\rm D} C_{\rm D} \left ( O^{\rm R}_{11} O^{\rm R}_{33} 
+ O^{\rm R}_{13} O^{\rm R}_{31} \right ) \right ] \sin\sigma \; .
\end{eqnarray}
Combining Eqs. (13) and (16), one can clearly see that the CP-violating
asymmetry $\varepsilon_1$ is proportional to $A_{\rm D}|B_{\rm D}|
\sin\sigma$. Hence $\sigma$ is the only source of CP violation for the
decays of heavy right-handed neutrinos in our ansatz.

Apart from $\sin\beta_{\rm susy}$ or $\tan\beta_{\rm susy}$, 
the free parameters of $\varepsilon_1$ 
include $m_i$, $M_i$ (for $i=1,2,3$) and $\sigma$. Current neutrino
oscillation data allow us to constrain $m_i$ or equivalently 
$x_\nu$, $z_\nu$ and $m_3$ to an acceptable degree of 
accuracy \cite{XZ}. The ratio
\begin{equation}
r \; \equiv \; \frac{B_{\rm R}}{C_{\rm R}} \; = \;
\frac{B_\nu}{C_\nu} \; = \; \sqrt{\frac{\left (1 -x_\nu \right )
\left (1 + z_\nu \right ) \left (x_\nu -z_\nu \right )}
{x_\nu z_\nu}} \; 
\end{equation}
can then be determined. Note that Eq. (17) remains valid, if 
$(x_\nu, z_\nu)$ are replaced by $(x^{~}_{\rm R}, z^{~}_{\rm R})$.
This implies that $x^{~}_{\rm R}$ and $z^{~}_{\rm R}$ are correlated
with each other for a given value of $r$. Indeed, it is straightforward
to obtain
\begin{equation}
z^{~}_{\rm R} = \frac{\displaystyle \sqrt{\left [ 
\left (1-x^{~}_{\rm R} \right )^2
+ r^2 x^{~}_{\rm R} \right ]^2 + 4 x^{~}_{\rm R} \left ( 1-x^{~}_{\rm R}
\right )^2} - \left [ \left ( 1-x^{~}_{\rm R} \right )^2 +
r^2 x^{~}_{\rm R} \right ]}{2 \left (1 - x^{~}_{\rm R} \right )} \; .
\end{equation}
Taking account of Eq. (18), we are left with four unknown parameters
to evaluate the magnitude of $\varepsilon_1$; namely,
$M_1$, $x^{~}_{\rm R}$, $\sigma$ and $\tan\beta_{\rm susy}$. 

In the spirit of thermal leptogenesis \cite{FY}, the CP-violating 
asymmetry $\varepsilon_1$ may lead to a net lepton number asymmetry 
$Y_{\rm L} \equiv (n^{~}_{\rm L} - n^{~}_{\rm\bar L})/{\bf s}
= \varepsilon_1 d/g^{~}_*$, where $g^{~}_* = 228.75$ 
is an effective number characterizing the relativistic degrees of freedom 
which contribute to the entropy {\bf s} of the early universe, and $d$ 
accounts for the dilution effects induced by the lepton-number-violating 
wash-out processes. This lepton number asymmetry  
is eventually converted into a net baryon number asymmetry 
$Y_{\rm B} \equiv (n^{~}_{\rm B} - n^{~}_{\rm\bar B})/{\bf s}$ via
nonperturbative sphaleron processes \cite{Kuzmin}
\footnote{We are grateful to Yanagida for clarifying our misunderstanding 
of Ref. \cite{Turner} and calling our particular attention 
to Hamaguchi's PhD thesis \cite{H}, in which the relationship
between $Y_{\rm B}$(final) and $Y_{\rm L}$(initial) is clearly discussed.}:
$Y_{\rm B} \approx -0.35 Y_{\rm L}$ in the MSSM with
three fermion families and two Higgs doublets. Although the dilution
factor $d$ can be computed by solving the full set of Boltzmann
equations, it is more convenient to adopt a simple analytical 
approximation of $d$ proposed in Ref. \cite{d}:
$d \approx 0.02 \times \left (0.01 ~{\rm eV}/\tilde{m}_1 \right )^{1.1}$,
where $\tilde{m}_1 \equiv (\tilde{M}_{\rm D}^\dagger 
\tilde{M}_{\rm D})_{11}/M_1$ is an effective neutrino mass 
parameter and its plausible magnitude is expected to lie in the range 
$10^{-2} ~ {\rm eV} \lesssim \tilde{m}^{~}_1 \lesssim 1 ~ {\rm eV}$
(the so-called strong washout regime \cite{d}). If our phenomenological
ansatz of lepton mass matrices is viable for leptogenesis, it should be 
able to reproduce the observed magnitude of $Y_{\rm B}$ (i.e., 
$7 \times 10^{-11} \lesssim Y_{\rm B} \lesssim 10^{-10}$ \cite{WMAP}).

To evaluate $\varepsilon_1$ and $Y_{\rm B}$, we adopt 
$x_\nu \approx 1/3$, $z_\nu \approx 1/12$ and $m_3 \approx 0.05$ eV
as favorable inputs given in Ref. \cite{XZ}. In addition, we typically
take $\tan\beta_{\rm susy} \approx 10$ and $x^{~}_{\rm R} \approx 1/4$.
We allow $M_1$ and $\sigma$ to vary, in order to reproduce $Y_{\rm B}$
in its afore-mentioned range. Then we arrive at the parameter space of
$M_1$ and $\sigma$, as shown in Fig. 1. The lower bound of $M_1$ is
about $2.5\times 10^{10}$ GeV, while the lower limit of $\sigma$ 
approximately reads $8.0^\circ$. When $M_1$ is much higher than
$10^{12}$ GeV, the value of $\sigma$ becomes closer to $57^\circ$.
Note that there is in general a potential conflict between achieving
successful thermal leptogenesis and avoiding overproduction of
gravitinos in the conventional seesaw model with supersymmetry \cite{Raidal}. 
If the mass scale of gravitinos is of ${\cal O}(1)$ TeV, one must
have $M_1 \lesssim 10^8$ GeV. This limit is apparently disfavored in
our ansatz. Such a problem could be circumvented in other supersymmetric
breaking mediation scenarios (e.g., gauge mediation \cite{Ross}) or in
a class of supersymmetric axion models \cite{AY}, where the
gravitino mass can be much lighter in spite of the very high reheating 
temperature.

It is worth pointing out that our ansatz can clearly be distinguished
from the interesting Fukugita-Tanimoto-Yanagida (FTY) ansatz of lepton 
mass matrices \cite{FTY} by comparing between their consequences on
leptogenesis. In the FTY ansatz, $M_l$ and $M_{\rm D}$ are also of the 
Fritzsch texture, but $M_{\rm R} = M_0 {\bf 1}$ with $\bf 1$ being the 
unit matrix has been assumed. Hence the texture of $M_\nu$ is 
different from that of $M_l$ or $M_{\rm D}$. The exact mass degeneracy 
of three right-handed neutrinos (i.e.,$M_i = M_0$ for $i=1,2,3$) 
immediately implies that the rephasing invariant of CP violation 
\begin{eqnarray}
\Delta_{\rm CP} & \equiv & {\rm Im} {\rm Tr}
\left [ Y^\dagger_\nu Y_\nu M^\dagger_{\rm R} M_{\rm R} M^\dagger_{\rm R}
Y^T_\nu Y^*_\nu M_{\rm R} \right ]
\nonumber \\
& = & \frac{1}{\langle H_2\rangle^4} 
\sum_{i<j} \left \{ M_i M_j \left (M^2_j - M^2_i \right )
{\rm Im} \left [(\tilde{M}^\dagger_{\rm D} \tilde{M}_{\rm D})_{ij} 
\right ]^2 \right \} \; ,   
\end{eqnarray}
which is relevant for leptogenesis \cite{BP}, is actually vanishing.
We can therefore conclude that there will be no lepton number asymmetry 
between $N_1 \rightarrow l + H^{\rm c}_2$ and 
$N_1 \rightarrow l^{\rm c} + H_2$ decays in the FTY ansatz, although
it {\it does} allow leptonic CP violation to show up in low-energy 
neutrino oscillations. In order
to accommodate leptogenesis, a slight modification of the FTY hypothesis
is necessary. The simplest way might be to assume that three heavy
right-handed neutrinos have a near mass degeneracy \cite{XZ05}, such 
that the cosmological baryon number asymmetry may arise from the 
resonantly-enhanced thermal leptogenesis \cite{P}.

\section{Lepton flavor violation}

We proceed to discuss the consequence of our phenomenological scenario
on lepton flavor violation in the framework of MSSM with three heavy
right-handed neutrinos. 
We restrict ourselves to a very simple and conservative case where 
supersymmetry is broken in a hidden sector and the breaking is 
transmitted to the observable sector by a flavor blind mechanism,
such as gravity \cite{Ross}. Then all the soft breaking terms are 
diagonal at high energy scales, and the only source of lepton flavor 
violation in the charged lepton sector is the radiative correction
to the soft terms through the neutrino Yukawa couplings. In other
words, the low-energy lepton-flavor-violating processes 
$l_j \rightarrow l_i \gamma$ (for $i,j = e, \mu, \tau$ and 
$m^{~}_j > m^{~}_i$) are induced by the renormalization-group
effects of the slepton mixing. 
The off-diagonal elements of the left-handed slepton mass matrix 
can be written as \cite{Ellis}
\begin{equation}
\left (M^2_{\tilde L} \right )_{ij} \approx 
- \frac{3 m^2_0 + A^2_0}{8 \pi^2 v^2 \sin^2\beta_{\rm susy}}
\left ({\tilde M}_{\rm D} L {{\tilde M}_{\rm D}}^\dagger \right )_{ij}
\end{equation}
in the leading-logarithmic approximation,
where $m_0$ and $A_0$ denote the universal scalar soft mass and 
the trilinear term at the GUT scale, respectively. Note that
the diagonal matrix $L$ in Eq. (19) measures the difference between 
the scale of heavy Majorana neutrinos and that of GUT, 
\begin{equation}
L \; = \; \left ( \matrix{
\ln \frac{M_{\rm GUT}}{M_1} & 0 & 0 \cr
0 & \ln \frac{M_{\rm GUT}}{M_2} & 0 \cr
0 & 0 & \ln \frac{M_{\rm GUT}}{M_3} \cr} \right ) \; .
\end{equation}
The branching ratios of $l_j \rightarrow l_i \gamma$ can approximately
be given by
\begin{equation}
{\rm Br}(l_j \rightarrow l_i \gamma) \approx 
\frac{\alpha^3}{G^2_{\rm F}} \cdot \frac{|(M^2_{\tilde L})_{ij}|^2}
{m^8_{\rm S}} \tan^2\beta_{\rm susy} 
\end{equation}
with $m^8_{\rm S} \approx 0.5m^2_0 M^2_{1/2} (m^2_0 + 0.6M^2_{1/2})^2$,
where $M_{1/2}$ denotes the gaugino mass \cite{Tanimoto}. Once the
texture of $\tilde{M}_{\rm D}$ is specified, it will be possible to
get concrete predicitions for ${\rm Br}(l_j \rightarrow l_i \gamma)$.  

Given the Fritzsch-like textures of $M_a$ (for $a=l,\nu, {\rm D}, {\rm R}$),
the quantity ${\tilde M}_{\rm D} L {{\tilde M}_{\rm D}}^\dagger$   
can be rewritten as 
\begin{equation}
{\tilde M}_{\rm D} L {{\tilde M}_{\rm D}}^\dagger \; = \; 
\left (Q O^T_l P'{\overline M}_{\rm D} P O_{\rm R} \right )
L \left (Q O^T_l P'{\overline M}_{\rm D} P O_{\rm R} \right )^{\dagger}
\end{equation}
in the chosen flavor basis (i.e., both $M_l$ and $M_{\rm R}$ are 
diagonal, real and positive), where 
$P' \equiv P^\dagger_l P_{\rm D} = {\rm Diag}\{e^{i\alpha}, e^{i\rho}, 1\}$ 
with 
$\alpha \equiv (\varphi^{~}_{\rm D}-\phi^{~}_{\rm D})- 
(\varphi^{~}_l-\phi^{~}_l)$
and $\rho \equiv \phi^{~}_{\rm D}-\phi^{~}_l$,
and all the other matrices have been given
before. Note that the phase $\alpha$ defined here is just the
one defined in Eq. (12) for the lepton flavor mixing matrix $V$,
because the phase equation
$\varphi^{~}_{\rm D}-\phi^{~}_{\rm D}=\varphi_{\nu}-\phi_{\nu}=
\varphi^{~}_{\rm R}-\phi^{~}_{\rm R}$ {\it does} hold in our ansatz.
Furthermore, it is easy to prove
\begin{equation}
\rho + \sigma = 2\phi^{~}_{\rm D} - \phi^{~}_{\rm R} - \phi^{~}_l
= \phi^{~}_\nu - \phi^{~}_l = \beta \; ,
\end{equation}
where the seesaw phase relation $\phi^{~}_\nu = 
2\phi^{~}_{\rm D}-\phi^{~}_{\rm R}$ has been used. Let us summarize 
the phase parameters appearing in three categories of phenomena
at this point:
\begin{eqnarray}
&& {\rm Neutrino ~ oscillations:} ~~ \alpha ~ {\rm and} ~ \beta \; ;
\nonumber \\
&& {\rm Thermal ~ leptogenesis:} ~~ \sigma \; ;
\nonumber \\
&& {\rm Lepton ~ flavor ~ violation:} ~~ \alpha ~ {\rm and} ~ \rho = 
\beta - \sigma \; .
\nonumber 
\end{eqnarray}    
We see that $V$, $Y_{\rm B}$ and ${\rm Br}(l_j \rightarrow l_i\gamma)$
totally involve three free phases. Among them, $\alpha$ and $\beta$ can 
be determined from the precise measurement of lepton flavor mixing and
CP violation in neutrino oscillations. Current neutrino oscillation
data {\it do} favor $\beta \approx \pi$, but they are unable to provide 
a narrow constraint on $\alpha$ \cite{XZ}. If the value of $\alpha$ is
fixed, nevertheless, one may examine the dependence of  
${\rm Br}(l_j \rightarrow l_i\gamma)$ on $\sigma$ 
with the help of $\rho \approx \pi - \sigma$. 
 
To illustrate, we take $\alpha \approx 30^\circ$,
$\beta \approx 180^\circ$ and $x^{~}_{\rm R} \approx 1/4$ in addition 
to $x_\nu \approx 1/3$, $z_\nu \approx 1/12$ and 
$m_3 \approx 0.05$ eV \cite{XZ}. 
The supersymmetric parameters relevant to our calculation are typically 
chosen as $\tan\beta_{\rm susy} \approx 10$, $A_0 \approx 0$,
$m^{~}_0 \approx 100$ GeV, $M_{1/2} \approx 300$ GeV and
$M_{\rm GUT} \approx 2 \times 10^{16}$ GeV. Then we are left with 
two free parameters $M_1$ and $\sigma$, just like the case of thermal
leptogenesis. Allowing $M_1$ and $\sigma$ to vary, we calculate 
${\rm Br}(l_j \rightarrow l_i\gamma)$ by using Eqs. (19)--(23) and by
including the leptogenesis constraint 
$7 \times 10^{-11} \lesssim Y_{\rm B} \lesssim 10^{-10}$. The 
experimental upper bounds of ${\rm Br}(l_j \rightarrow l_i\gamma)$
should certainly be taken into account \cite{Aoki}: 
\begin{eqnarray}
{\rm Br}(\mu \rightarrow e\gamma) & < & 1.2 \times 10^{-11} \; ,
\nonumber \\
{\rm Br}(\tau \rightarrow e\gamma) & < & 3.6 \times 10^{-7} \; ,
\nonumber \\
{\rm Br}(\tau \rightarrow \mu \gamma) & < & 3.1 \times 10^{-7} \; .
\end{eqnarray}
Our numerical results are shown in Figs. 2--4. Once can see that
the predicted branching ratios of $\mu \rightarrow e\gamma$,
$\tau \rightarrow e\gamma$ and $\tau \rightarrow \mu \gamma$ are
all below their corresponding experimental upper limits. A
generic feature of ${\rm Br}(l_j \rightarrow l_i\gamma)$ is that
they increase with the phase parameter $\sigma$. In particular,
${\rm Br}(\mu \rightarrow e\gamma) \sim 10^{-12}$ is reachable for 
$\sigma \sim 56^\circ$. Of course, the magnitudes of 
${\rm Br}(l_j \rightarrow l_i\gamma)$ will change, if different
values of the supersymmetric parameters are input. Instead of 
examining the full parameter space, here we have paid our main
attention to the parameter correlation between leptogenesis and
lepton-flavor-violating processes.

Let us remark that the gravity mediation scenario used
in evaluating lepton flavor violation is in potential conflict
with the lower bound of $M_1$ obtained from thermal leptogenesis 
in our ansatz. This problem, usually referred to as the gravitino 
problem, exists in many supersymmetric seesaw models (see, e.g.,
Ref. \cite{Raidal} and references therein). One possible way out 
is to fine-tune the model parameters (i.e., those of $Y_\nu$ and
$M_{\rm R}$) or assume the masses of heavy right-handed
neutrinos to be nearly degenerate \cite{P} within the thermal
leptogenesis mechanism. As for our universal Fritzsch-like textures 
of lepton mass matrices, however, we are not left with much room 
for the fine-tuning of model parameters. The assumption of a near 
mass degeneracy for heavy Majorana neutrinos seems not to be natural
either, although it is not impossible. 

We find that there actually exists an interesting solution to the 
gravitino problem \cite{AY}, which is in no apparent conflict with
our phenomenological scenario. Its key point is that the axino and
gravitino can be the lightest (of ${\cal O}(1)$ keV) and the next
lightest (of ${\cal O}(10^2)$ GeV) supersymmetric particles,
respectively, in a large class of supersymmetric axion models. Thus
the reheating temperature is allowed to reach $10^{15}$ GeV or 
so \cite{AY}, high enough for the thermal leptogenesis mechanism to 
work. Some more discussions in this connection will be presented
elsewhere \cite{XZ05}, since they are already beyond the scope of 
the present work. 

\section{Summary}

We have incorporated the seesaw mechanism with six Fritzsch-like 
textures of lepton mass matrices. It is found that the seesaw
relation holds under a particular condition, and the consequences 
of those textures on lepton flavor mixing are exactly the same. 
Applying this simple ansatz to thermal leptogenesis,
we have shown that CP violation in the lepton-number-violating 
decays of heavy right-handed neutrinos
depends only upon a single phase parameter and the cosmological 
baryon number asymmetry can naturally be explained. The 
lepton-flavor-violating processes $\mu \rightarrow e \gamma$,
$\tau \rightarrow \mu \gamma$ and $\tau \rightarrow e \gamma$
have also been calculated. An interesting result is that the
branching ratios of those rare processes rely strongly upon the 
phase parameters appearing in leptogenesis and in neutrino 
oscillations. We expect that a stringent test of our 
phenomenological scenario will be available in the near future,
when more precise experimental data are accumulated.

\acknowledgments{ 
We are deeply indebted to T. Yanagida for bringing Refs. \cite{H,AY}
to our attention and clarifying our
misunderstanding of the sphaleron and gravitino problems. We
are also grateful to W.L. Guo and J.W. Mei for helpful discussions. 
This work was supported in part by the National Natural Science
Foundation of China.}

\newpage

\begin{table}
\caption{The Fritzsch-like lepton mass matrix $M_a$ 
(for $a = l, {\rm D}, {\rm R}, \nu$), where $\arg ({\cal A}_a) \equiv 0$, 
$\arg ({\cal B}_a) \equiv \phi^{~}_a$ and 
$\arg ({\cal C}_a) \equiv \varphi^{~}_a$.
The phase matrix $P_a$ and the real orthogonal matrix $O_a$ 
are defined to diagonalize $M_a$ via the transformation 
$(P^*_a O_a Q)^T M_a (P^*_a O_a Q)$ with 
$Q \equiv {\rm Diag} \{1, i, 1\}$.
The explicit expressions of $O^a_{ij}$ (for $i,j =1,2,3$) are
given in Eq. (9).}
\begin{center}
\begin{tabular}{c|ccc}
Pattern & $M_a$ & $P_a$ & $O_a$  \\ \hline 
(A) &
$\left ( \matrix{
{\bf 0} & {\cal C}_a & {\bf 0} \cr
{\cal C}_a & {\bf 0} & {\cal B}_a \cr
{\bf 0} & {\cal B}_a & {\cal A}_a \cr} \right )$ &
$\left ( \matrix{
e^{i(\varphi^{~}_a -\phi^{~}_a)} & 0 & 0 \cr
0 & e^{i\phi^{~}_a} & 0 \cr
0 & 0 & 1 \cr} \right )$ &
$\left ( \matrix{
O^a_{11} & O^a_{12} & O^a_{13} \cr
O^a_{21} & O^a_{22} & O^a_{23} \cr
O^a_{31} & O^a_{32} & O^a_{33} \cr} \right )$ 
\\ \hline 
(B) &
$\left ( \matrix{
{\bf 0} & {\bf 0} & {\cal C}_a \cr
{\bf 0} & {\cal A}_a & {\cal B}_a \cr
{\cal C}_a & {\cal B}_a & {\bf 0} \cr} \right )$ &
$\left ( \matrix{
e^{i(\varphi^{~}_a -\phi^{~}_a)} & 0 & 0 \cr
0 & 1 & 0 \cr
0 & 0 & e^{i\phi^{~}_a} \cr} \right )$ &
$\left ( \matrix{
O^a_{11} & O^a_{12} & O^a_{13} \cr
O^a_{31} & O^a_{32} & O^a_{33} \cr
O^a_{21} & O^a_{22} & O^a_{23} \cr} \right )$ 
\\ \hline 
(C) &
$\left ( \matrix{
{\bf 0} & {\cal C}_a & {\cal B}_a \cr
{\cal C}_a & {\bf 0} & {\bf 0} \cr
{\cal B}_a & {\bf 0} & {\cal A}_a \cr} \right )$ &
$\left ( \matrix{
e^{i\phi^{~}_a} & 0 & 0 \cr
0 & e^{i(\varphi^{~}_a-\phi^{~}_a)} & 0 \cr
0 & 0 & 1 \cr} \right )$ &
$\left ( \matrix{
O^a_{21} & O^a_{22} & O^a_{23} \cr
O^a_{11} & O^a_{12} & O^a_{13} \cr
O^a_{31} & O^a_{32} & O^a_{33} \cr} \right )$ 
\\ \hline 
(D) &
$\left ( \matrix{
{\bf 0} & {\cal B}_a & {\cal C}_a \cr
{\cal B}_a & {\cal A}_a & {\bf 0} \cr
{\cal C}_a & {\bf 0} & {\bf 0} \cr} \right )$ &
$\left ( \matrix{
e^{i\phi^{~}_a} & 0 & 0 \cr
0 & 1 & 0 \cr
0 & 0 & e^{i(\varphi^{~}_a-\phi^{~}_a)} \cr} \right )$ &
$\left ( \matrix{
O^a_{21} & O^a_{22} & O^a_{23} \cr
O^a_{31} & O^a_{32} & O^a_{33} \cr
O^a_{11} & O^a_{12} & O^a_{13} \cr} \right )$ 
\\ \hline 
(E) &
$\left ( \matrix{
{\cal A}_a & {\bf 0} & {\cal B}_a \cr
{\bf 0} & {\bf 0} & {\cal C}_a \cr
{\cal B}_a & {\cal C}_a & {\bf 0} \cr} \right )$ &
$\left ( \matrix{
1 & 0 & 0 \cr
0 & e^{i(\varphi^{~}_a-\phi^{~}_a)} & 0 \cr
0 & 0 & e^{i\phi^{~}_a} \cr} \right )$ &
$\left ( \matrix{
O^a_{31} & O^a_{32} & O^a_{33} \cr
O^a_{11} & O^a_{12} & O^a_{13} \cr
O^a_{21} & O^a_{22} & O^a_{23} \cr} \right )$ 
\\ \hline 
(F) &
$\left ( \matrix{
{\cal A}_a & {\cal B}_a & {\bf 0} \cr
{\cal B}_a & {\bf 0} & {\cal C}_a \cr
{\bf 0} & {\cal C}_a & {\bf 0} \cr} \right )$ &
$\left ( \matrix{
1 & 0 & 0 \cr
0 & e^{i\phi^{~}_a} & 0 \cr
0 & 0 & e^{i(\varphi^{~}_a-\phi^{~}_a)} \cr} \right )$ &
$\left ( \matrix{
O^a_{31} & O^a_{32} & O^a_{33} \cr
O^a_{21} & O^a_{22} & O^a_{23} \cr
O^a_{11} & O^a_{12} & O^a_{13} \cr} \right )$ 
\\ 
\end{tabular}
\end{center}
\end{table}

\newpage

\begin{figure}[t]
\vspace{0cm}
\epsfig{file=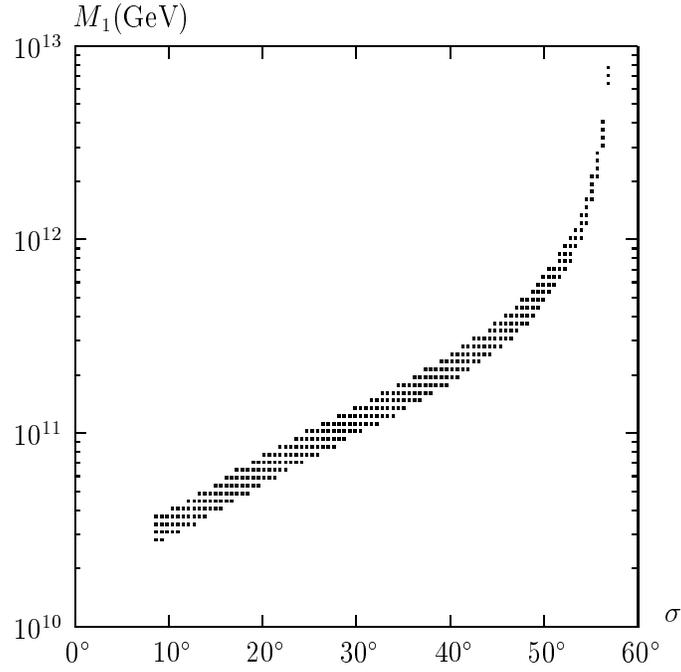,bbllx=1.5cm,bblly=7cm,bburx=17cm,bbury=28cm,%
width=14cm,height=20cm,angle=0,clip=0}
\vspace{-8.0cm}
\caption{The allowed ranges of $\sigma$ and $M_1$ to reproduce
$7 \times 10^{-11} \leq Y_{\rm B} \leq 10^{-10}$ via leptogenesis, 
where $m^{~}_3 \approx 0.05$ eV, $x_\nu \approx 1/3$, $z_\nu \approx 1/12$, 
$x^{~}_{\rm R} \approx 1/4$ and $\tan\beta_{\rm susy} \approx 10$ have 
typically been input.}
\end{figure}

\begin{figure}[t]
\vspace{0cm}
\epsfig{file=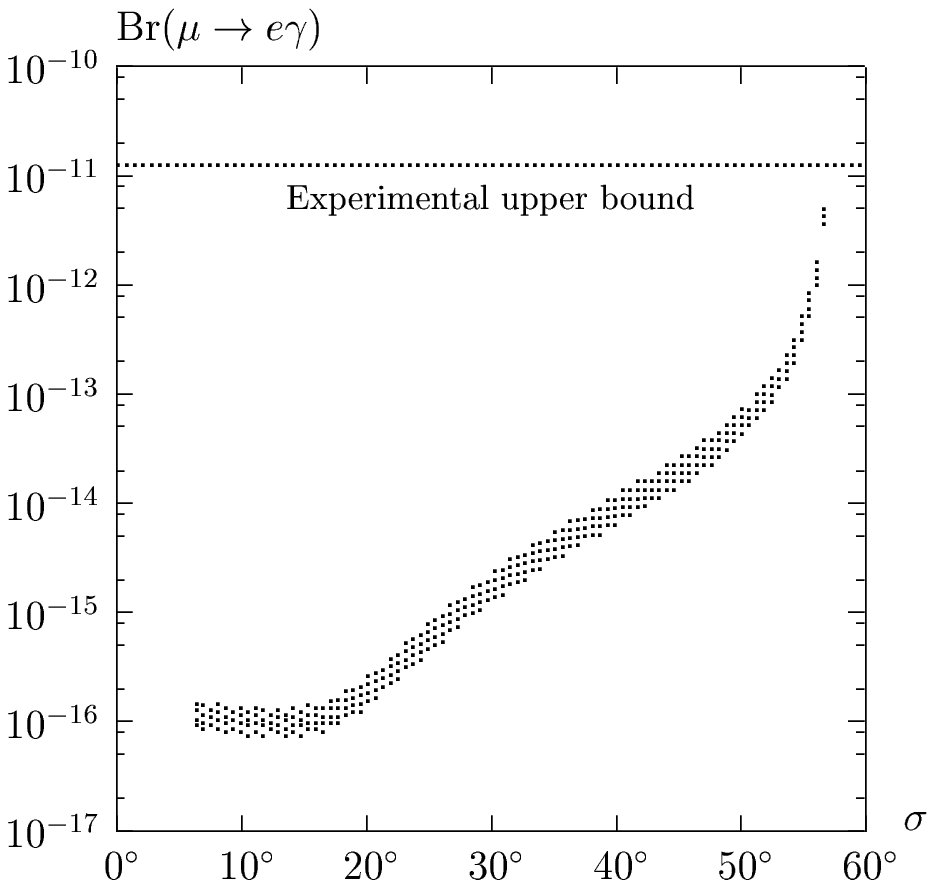,bbllx=1.5cm,bblly=7cm,bburx=17cm,bbury=28cm,%
width=14cm,height=20cm,angle=0,clip=0}
\vspace{-8.3cm}
\caption{The allowed ranges of $\sigma$ and 
${\rm Br}(\mu \rightarrow e\gamma)$, where 
$m^{~}_3 \approx 0.05$ eV, $x_\nu \approx 1/3$, $z_\nu \approx 1/12$, 
$x^{~}_{\rm R} \approx 1/4$ and $\tan\beta_{\rm susy} \approx 10$ have 
typically been input.}
\end{figure}

\begin{figure}[t]
\vspace{0cm}
\epsfig{file=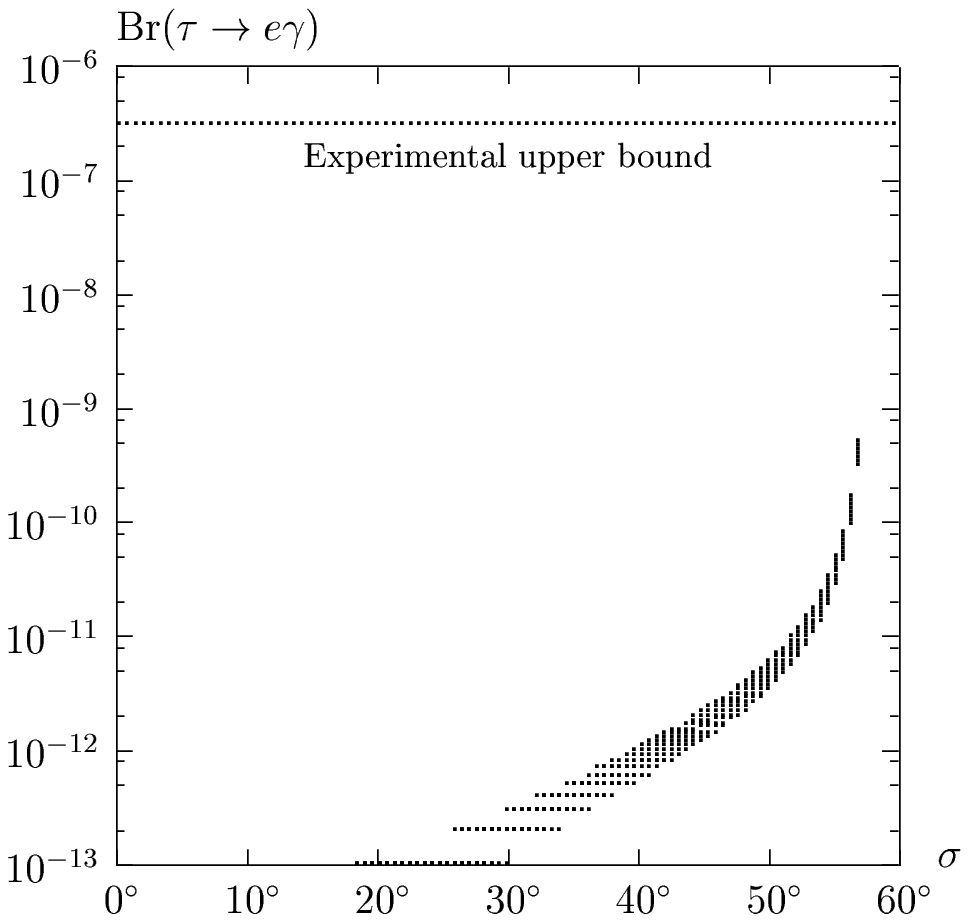,bbllx=1.5cm,bblly=7cm,bburx=17cm,bbury=28cm,%
width=14cm,height=20cm,angle=0,clip=0}
\vspace{-8cm}
\caption{The allowed ranges of $\sigma$ and 
${\rm Br}(\tau \rightarrow e\gamma)$, where
$m^{~}_3 \approx 0.05$ eV, $x_\nu \approx 1/3$, $z_\nu \approx 1/12$, 
$x^{~}_{\rm R} \approx 1/4$ and $\tan\beta_{\rm susy} \approx 10$ have 
typically been input.}
\end{figure}

\begin{figure}[t]
\vspace{0cm}
\epsfig{file=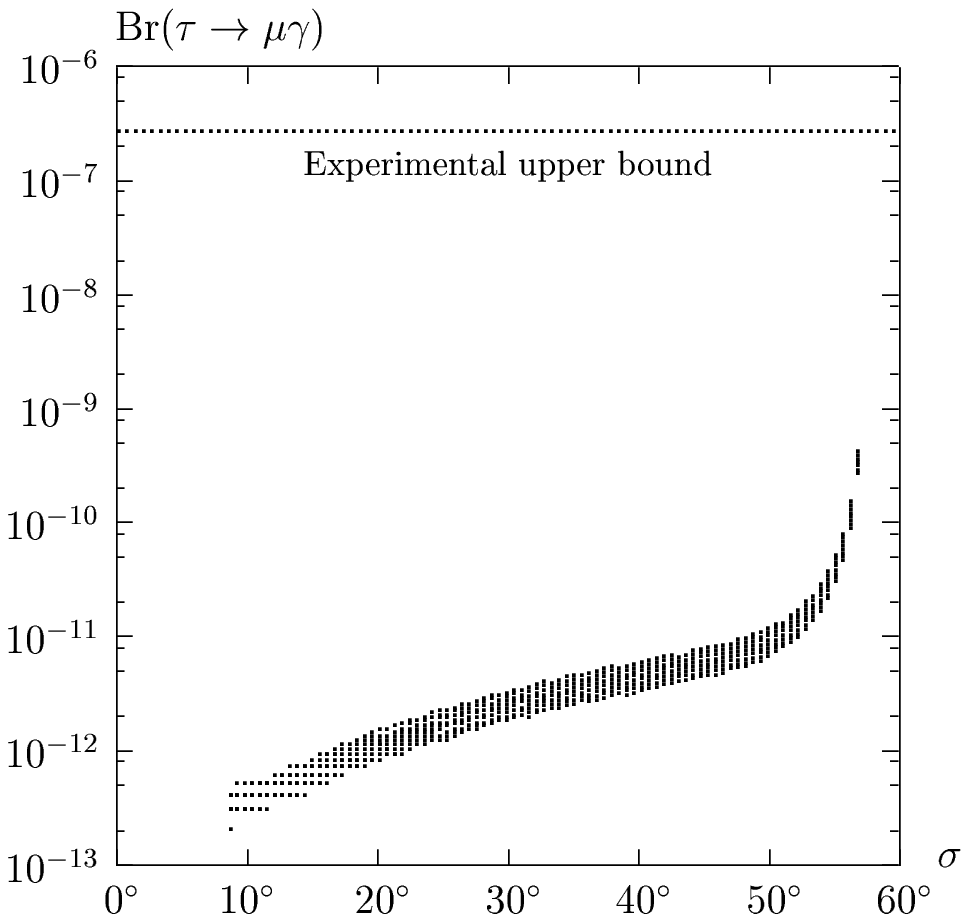,bbllx=1.5cm,bblly=7cm,bburx=17cm,bbury=28cm,%
width=14cm,height=20cm,angle=0,clip=0}
\vspace{-8cm}
\caption{The allowed ranges of $\sigma$ and 
${\rm Br}(\tau \rightarrow \mu \gamma)$, where
$m^{~}_3 \approx 0.05$ eV, $x_\nu \approx 1/3$, $z_\nu \approx 1/12$, 
$x^{~}_{\rm R} \approx 1/4$ and $\tan\beta_{\rm susy} \approx 10$ have 
typically been input.}
\end{figure}

\end{document}